\documentstyle[11pt]{article}    
\title{\bf Geometrical Origin of Tricritical Points of
various U(1)
Lattice Models
\thanks{Supported in part by Deutsche
Forschungsgemeinschaft under grant no. Kl.256.}
\thanks{Lecture presented by W. Janke.}}
\author{ {\it W. Janke} and {\it H. Kleinert}\\
        Institut f\"{u}r Theorie der Elementarteilchen\\
        Freie Universit\"{a}t Berlin\\
        Arnimallee 14\\
        D-1000 Berlin 33\\
        Germany}
\date {  }         
\begin{document}
\large
  \maketitle
{\Large
\begin{abstract}
We review the dual relationship between various compact U(1) lattice
models and Abelian Higgs models, the latter being the disorder field
theories of line-like topological excitations in the systems. We
point out that the predicted first-order transitions
in the Abelian Higgs models (Coleman-Weinberg mechanism) are,
in three dimensions,
in contradiction with direct numerical investigations in the
compact U(1) formulation since these yield continuous transitions
in the major part of the phase diagram.
In four dimensions, there are indications from Monte Carlo data
for a similar situation.
Concentrating on the strong-coupling
expansion in terms of geometrical objects, surfaces
or lines, with certain statistical weights, we
present semi-quantitative arguments explaining the observed cross-over
from first-order to continuous transitions by the balance
between the lowest two weights (``2:1 ratio'') of these geometrical objects.
\end{abstract}}
  \newpage
  \section{Introduction}
  \pagenumbering{arabic}
In many physical systems, even the
qualitative features along a line of phase transitions may
depend on the choice of the coupling constants which
parametrize the bare interaction energies. Quite often one
encounters the case where a continuous higher order
transition changes into a discontinuous first-order one. In
most cases, the cross-over behaviour is
characterized by a tricritical point (TCP)\cite{TCP}. In
field theoretic language, the origin of a TCP is easy
to understand on the basis of a Landau expansion of the
effective potential
in terms of
a order- (or disorder-) parameter. One source for a
TCP is the sign-change in the quartic term (assuming a
stabilizing sextic term) in some range of the parameter
space. A positive quartic term  leads to a second-order
transition, and a negative one to a first-order transition. A
vanishing quartic term is associated with the TCP. Another
important mechanism arises from the presence of a cubic
term. Such a cubic term may be generated,
in some range of the parameter space, by fluctuations
of other fields. This always drives the transition
first-order. While the analysis of the Landau expansion
is very elementary, the
non-trivial problem is, of course, its derivation
from the underlying physical system with
quite complicated microscopic forces (the most prominent
example being the Gorkov derivation \cite{gorkov} of the effective
Ginzburg-Landau theory of superconductivity). It is
therefore desirable to find alternative properties of the system which
allow at least for a semi-quantitative
understanding of the appearance of
TCP's. In this lecture we exhibit such properties for compact U(1)
lattice models, having in mind quite different physical
systems (D=dimension):

1. D=4 Quantumelectrodynamics (U(1)-LGT)

2. D=3 Superfluid Helium (XY-model)

3. D=3 Defect models of melting

The common point of these systems is that they can be studied in

a)
the weak-coupling or ``defect'' expansion. This is formulated
in terms of lines\footnote
{When the dimension is reduced by one, $D \rightarrow
D-1$, ``lines'' have to be replaced by ``points''.}
with long-range Biot-Savart-like interactions. This is why all three
systems can be described, alternatively, by Abelian Higgs models
in which the complex fields account for the line-like disorder of the systems
and the gauge fields for the long-range interactions \cite{book}.

b) The
strong-coupling or ``stress'' expansion. This is also formulated in
terms of geometrical objects (surfaces, lines), but with
no long-range interactions. This simplifies their
statistical behaviour and will provide for the desired alternative
criterion for the existence of TCP's.

For the specific example
of U(1) lattice gauge theory the different representations,
being dual to each other, and their corresponding
field-theoretic descriptions are summarized in
fig.1. The diagrams for the other systems are
completely analogous.
  \begin{figure}
     \vspace{7.0cm}
     \caption[1]{Interrelation between the dual ``defect''
              (weak-coupling) and ``stress''
              (strong-coupling) representations of U(1)
              lattice gauge theory and their equivalent
field theoretic descriptions.}
   \end{figure}
  \section{The ``Defect'' Representation}
Let us first look at the weak-coupling expansion of the various
U(1) lattice models. They all contain topological excitations with
long-range interactions caused by the Nambu-Goldstone
bosons of the U(1) symmetry. In a three-dimensional solid, the
topological excitations are line-like defects, dislocations and
disclinations \cite{defect} , and the Nambu-Goldstone bosons
are elastic
distortions (phonons), leading to a stress-field around each
defect and to Biot-Savart-like interactions between defect
elements. In superfluid Helium in three dimensions, the ``defects''
are vortex-lines \cite{vortex} and the long-range forces are
caused by the superflow. In four-dimensional U(1)-LGT
finally, the ``defects'' are world lines of magnetic
monopoles \cite{monopole} and the forces are due to electromagnetism.
The upper right box in fig.1 symbolizes the
equivalent disorder field description of these ``defect'' lines
in terms of a complex field
$\Psi$ interacting with a gauge field $A_{\mu}$. In
superfluid Helium and U(1)-LGT, the disorder field theory of
line-like defects turns out \cite{hk3d,hk4d} to be an
Abelian Higgs model. In a solid, it is a more complicated field theory
of a similar type, involving two disorder fields, one for dislocation
and one for disclination lines \cite{book}.

Let us briefly recall the
physical content of the dual equivalence. From Feynman's
path-integral representation of quantum mechanics, we know
that the statistical mechanics of one
fluctuating line (``orbit'') corresponds to the quantum
mechanics of one particle. It is then easy to see that the
grand-canonical ensemble of fluctuating lines corresponds to
the quantum mechanics of a
many-particle system. This in turn is described most conveniently by
a second quantized
field theory.
The long-range Biot-Savart-like nature of the interactions
between line-elements is what permits their description by
an Abelian gauge-field
$A_{\mu}$. The minimal coupling of $A_{\mu}$
to the disorder field $\Psi$ leads then immediately to an
Abelian Higgs model (scalar electrodynamics).
Indeed, in the lattice formulation of
interacting defect lines, these steps can be carried out
rigorously \cite{hk3d,hk4d}, leading to the Higgs action
\begin{equation}
{\cal A} = \int d^{D}\!x \left[ \frac{1}{4}F_{\mu\nu}^{2} + \frac{1}{2}
|(\partial
_{\mu} - ieA_{\mu}) \Psi|^{2} + \frac{1}{2} m^{2} |\Psi|^{2} +
\frac{1}{4} g |\Psi|^{4} + \ldots \right]
\label{eq:1}
\end{equation}
where $F_{\mu\nu} = \partial _{\mu} A_{\nu} - \partial _{\nu}
A_{\mu}$, and $e,m^{2},g>0$ can in principle be calculated from
the couplings of
the original compact U(1) lattice model, indicated by the circle in
the center of fig.1. According to the famous
Coleman-Weinberg argument \cite{coleman,halperin},
this Abelian Higgs model should
always undergo a first-order transition when the mass
parameter turns negative.
The argument goes
as follows: Assuming $\Psi = $ const., the gauge fields can
be integrated out, yielding a one-loop correction to (\ref{eq:1})
\begin{equation}
{\rm tr} \log (k^{2} + e^{2}|\Psi|^{2})
\label{eq:1.1}
\end{equation}
This amounts to additional $\Psi$ interactions of the
form
\begin{equation}
{\rm ~~}-|\Psi|^{3} {\rm ~~~~~~~~~~in~D=3}
\label{eq:1.2}
\end{equation}
and
\begin{equation}
|\Psi|^{4} \log |\Psi|        {\rm  ~~~~~~~in~D=4}
\label{eq:1.3}
\end{equation}
In both cases, the additional term should drive the transition
first-order.

In three
dimensions, this conclusion is in clear contradiction to the
well established fact that the original D=3 XY model has a continuous
phase transition \cite{3dxy,dasgupta}. The crucial
point in the whole argument is, of course, the assumption of
an almost constant $\Psi$-field. If this assumption breaks
down, then also the generation of a cubic term (in D=3) is
no longer reliable and the transition may stay continuous
even with $A_{\mu}$-field fluctuations. The decisive parameter is
the ratio of the length-scales of
$A_{\mu}$- and $\Psi$- fluctuations,
\begin{equation}
\kappa = \frac{1}{\sqrt{2}} \frac{{\rm penetration~depth }(A_{\mu})}
{{\rm coherence~length }
(\Psi)}
\label{eq:1.4}
\end{equation}
In three dimensions, it was possible to show that there
exists a tricritical value of $\kappa \approx 1/\sqrt{2}$,
i.e. near the separation line between type-I and type-II
superconductivity \cite{hk3d}.
For small $\kappa$, the Coleman-Weinberg mechanism is
valid, leading to a first-order transition,
whereas for large $\kappa$, the transition stays second
order.

This led to the suggestion that also four-dimensional scalar
electrodynamics could have a TCP \cite{hk4db}. By the above duality
arguments, this would also hold for U(1)-LGT.
Indeed, the
first-order nature predicted by the Coleman-Weinberg mechanism
was at odds with early Monte Carlo
investigations of the dual U(1)-LGT which all claimed evidence
for a continuous transition and even reported estimates for
critical indices \cite{earlyu1}. More recent work favors the existence
of both transition regions, first-order {\em and} second-order, in
agreement with ref.\cite{hk4db}. Only one paper claims the validity
of the Coleman-Weinberg mechanism everywhere.
The present status
will be reviewed in more detail in the next section.

The goal of this lecture is to present semi-quantitative
explanations for these TCP's, working within the dual
strong-coupling expansions of the various U(1) models. This
has the advantage that, since the geometrical
objects appearing in this expansion have no long-range
interactions, subtleties of the type (\ref{eq:1.1})--(\ref{eq:1.3})
are absent.
  \section{The ``Stress'' Representation}
Let us start by recalling the definitions of the various
U(1) models mentioned in the introduction. The partition
functions
\begin{equation}
Z = \sum_{\{\gamma{\rm - conf.}\}} (\prod B)
\label{eq:2}
\end{equation}
are products of local Boltzmann factors which may be chosen
as
\begin{equation}
B = \left\{
\begin{array}{ll}
e^{\beta \cos \Theta} & {\rm Wilson}\\
e^{\beta \cos \Theta + \gamma \cos 2\Theta} & {\rm Mixed}\\
\sum_{n}e^{-\frac{\beta_{V}}{2}(\Theta - 2 \pi n)^{2}} & {\rm
Villain}
\end{array}
\right\} {\rm action}
\label{eq:3}
\end{equation}
with $\Theta$ standing short for
\begin{equation}
\Theta = \left\{
\begin{array}{ll}
\nabla_{i}\gamma_{j} - \nabla_{j}\gamma_{i} \equiv
      (\nabla_{i}\gamma_{j})^{A} & {\rm Gauge}\\
\nabla_{i}\gamma
& {\rm XY}\\
\nabla_{i}\gamma_{j} + \nabla_{j}\gamma_{i} \equiv
      (\nabla_{i}\gamma_{j})^{S} & {\rm Melting}
\end{array}
\right\} {\rm model}
\label{eq:4}
\end{equation}
The product in (\ref{eq:2}) runs over the plaquettes or
links of the lattice, and
$\nabla_{i}$ are the usual lattice derivatives
($\nabla_{i}f(\vec x) = f(\vec x + \vec i) - f(\vec x))$.
The $\gamma$-configurations are, in U(1)-LGT,
the euclidean
electromagnetic fields, in the case of superfluid
Helium, the phase angles of the condensate, and in the
melting model, the
atomic displacements.
The common important property of these models
is the U(1) invariance
\begin{equation}
\gamma \rightarrow \gamma + 2 \pi
\label{eq:5}
\end{equation}
suggesting a Fourier (``character'') expansion of the
Boltzmann factors
\begin{equation}
B(\Theta) = \sum_{b=-\infty}^{\infty} W_{b} e^{ib\Theta}
\label{eq:6}
\end{equation}
with Fourier coefficients
\begin{equation}
W_{b} =
\int_{-\pi}^{\pi}\frac{d\Theta}{2\pi}B(\Theta)e^{-ib\Theta}
\label{eq:7}
\end{equation}
Inserting (\ref{eq:6}) into the partition function and
summing over the $\gamma$-configurations, one obtains the
strong-coupling expansion $(\bar{W}_{b} \equiv
W_{b}/W_{0})$
\begin{equation}
Z = (\prod W_{0}) \sum_{\{b{\rm-conf.}\}} (\prod \bar{W}_{b})
\label{eq:8}
\end{equation}
The algebraic structure of $\Theta$ in eq.(\ref{eq:4}) leads
to conservation laws which constraint the admissible
$b$-configurations. It is most transparent to characterize
them in terms of geometrical objects:
\begin{equation}
\{b{\rm-conf.}\} = {\rm closed} \left\{
\begin{array}{ll}
{\rm surfaces} & {\rm Gauge}\\
{\rm lines} & {\rm XY}\\
{\rm complicated~lines} & {\rm Melting}
\end{array}
\right.
\label{eq:8.1}
\end{equation}
In the U(1)-LGT, the $b$-variables correspond to the
electromagnetic field strengths, in the superfluid to the
superfluid currents, and in the melting model to the
physical stress. This is why we call the expansion (\ref{eq:8})
generically the ``stress''-representation. The geometrical
objects in (\ref{eq:8.1}) are subject only to short-range
contact interactions.
Notice that the geometrical characterization of these
``stress''-graph configurations does not depend on the
dimension D. This is in contrast to the
``defect''-representation where the dimensionality of the
geometrical objects depends on the dimension in which
the duality transformation is performed.
  \section{Analytical and Numerical Results}
Let us now
analyze the ``stress'' representation (\ref{eq:8}) in some detail.
To each choice of action in (\ref{eq:3}) corresponds a
special ``natural'' set of weights $\bar {W}_{b}$ in
(\ref{eq:8}). Their relative importance with increasing
strength $b=\pm 1, \pm 2, \ldots$ can be studied
conveniently by simulating each model with different actions and
comparing their thermodynamic quantities such as their
internal energies.

We start with the comparison
between Wilson's and Villain's action for which the weights
are
$\bar{W}_{b}^{W} = I_{b}(\beta)/I_{0}(\beta)$ ($I_{b}:$ modified
Bessel function) and $\bar{W}_{b}^{V} =
\exp(-b^{2}/(2\beta_{V}))$, respectively. The two actions
are made as similar as possible by equating the
$b=\pm 1$ weights which amounts to relating the
Villain parameter $\beta_{V}$ to the Wilson parameter
$\beta$ as follows (see fig.2)
\begin{equation}
\beta_{V}(\beta) = -\frac{1}{2 \log
[I_{1}(\beta)/I_{0}(\beta)]}
\label{eq:9}
\end{equation}
This relation was first written down by Villain
\cite{villain} when analyzing the Wilson type of
action of the XY model in terms of the discrete Gaussian model
(Villain approximation).
If furthermore the overall normalizations of the partition
functions are adjusted, then the difference between the
two actions lies all in the higher weights,
$\bar{W}_{b}^{W}(\beta)=I_{b}(\beta)/I_{0}(\beta) \neq
[I_{1}(\beta)/I_{0}(\beta)]^{b}=\exp(-b^{2}/(2\beta_{V}))=
\bar{W}_{b}^{V}$ for $b \geq 2$.
  \begin{figure}
    \vspace{7.5cm}
    \caption[2]{The parameter $\beta_{V}$ of the Villain
approximation versus $\beta$ of the corresponding Wilson
action. The curve results from the requirement of equal
weights for graphs of
strength-1 in the ``stress'' representation for both
actions.}
   \end{figure}
In order to measure their importance,
we have performed Monte Carlo simulations in the
representation (\ref{eq:2})
with both Wilson's and Villains's action \cite{jkvill}. As a typical
example, we compare in fig.3 our results for the internal energy of
the U(1) lattice gauge model.
The excellent agreement for
low $\beta$ up to the phase transition around $\beta \approx
1$ demonstrates that, in this range, the systems are dominated
by $b=\pm1$ excitations.
   \begin{figure}
      \vspace{7.5cm}
      \caption[3]{The internal energy of D=4 U(1) lattice gauge model
with Wilson action in comparison with the Villain approximation.}
   \end{figure}
It is therefore not surprising that also the
transition temperatures of the Villain action (with
$\beta_{V}$ as a free parameter) are mapped by (\ref{eq:9})
very precisely onto the corresponding ones of the Wilson
action (see table 1). Furthermore, from fig.3 we read off
that at and
%
 \begin{table}
 \begin{center}
  \begin{tabular}{|c|c|c|c|}
   \hline 
    Model  &  $\beta_{Vc}$ &$\beta_{c}$(from VA) & $\beta_{c}$\\
   \hline
4D-U(1) & $0.643^{a}$ & 1.04 & $1.0111^{b}$ \\
3D-XY   & $0.33^{c}$  & 0.45 & $0.4539^{d}$ \\
2D-XY   & $0.73^{e}$  & 1.18 & $1.12^{f},1.18^{g},1.14^{h}$ \\ 
   \hline
  \end{tabular}
 \end{center}
 \caption[tab1]{Transition temperatures of U(1) lattice gauge model for 
D=4, and
of XY model for D=2,3 compared 
with the values obtained from the 
Villain mapping $\beta_{V} = 
-[2\log(I_{1}(\beta)/I_{0}(\beta))]^{-1}$. The values are 
taken from: a) Jers\'{a}k et al. \cite{jersak}, b) Gupta et 
al. \cite{gupta}, c) Dasgupta and Halperin \cite{dasgupta}, 
d) Ferer et al. \cite{3dxy}, e) Shugard et al. 
\cite{shugard}, f) Tobochnik and Chester \cite{tobochnik}, 
g) Samuel and Yee \cite{samuel}, h) Harten and Suranyi 
\cite{harten}.}
 \end{table}

above the phase transition the graphs of higher strengths
proliferate much stronger using the Wilson action.
Among these we expect the graphs of strength-2 to be most
important. This suggests that the weight ratio
$\bar{W}_{2}/\bar{W}_{1}$ should be the relevant distinguishing
feature of the different actions. In the sequel, it will be
called the ``2:1 ratio''. In the Wilson action, the ``2:1
ratio'' is much larger than in the Villain action as is
shown in table 2.

It is easy to convince ourselves that the
higher ratios $\bar{W}_{3}/\bar{W}_{1},\ldots$ are
practically irrelevant near the phase transition.
    \begin{table}
 \begin{center}
  \begin{tabular}{|c|c|c|}
   \hline 
   $\bar{W}_{2}/\bar{W}_{1}$  & XY($\beta=0.45$) &  Gauge($\beta=1.0$) \\
   \hline
   Wilson & 0.11 & 0.24 \\
   Villain& 0.01 & 0.10 \\
   \hline
  \end{tabular}
 \end{center}
 \caption{The ``2:1 ratio'' near the transition points of 
the D=3 XY and D=4 U(1) lattice gauge models. It shows that the 
``stress'' graphs of strength-2 are much more important in the 
Wilson than in the Villain action.}
 \end{table}

For this purpose we compare the Villain action
with further Monte Carlo simulations of the mixed
$\beta-\gamma$ action (see eq.(\ref{eq:3})), which is chosen
in such a way that the ``2:1 ratio'' coincides with that of
the Villain action. The coincidence is achieved by requiring
\begin{eqnarray}
\frac{I_{1}(\beta,\gamma)}{I_{0}(\beta,\gamma)} & = &
\exp(-\frac{1}{2\beta_{V}}) \label{eq:10a}\\
\frac{I_{2}(\beta,\gamma)}{I_{0}(\beta,\gamma)} & = &
\exp(-\frac{4}{2\beta_{V}})
\label{eq:10b}
\end{eqnarray}
where $I_{b}(\beta,\gamma)$ are generalized modified Bessel
functions
\begin{equation}
I_{b}(\beta,\gamma) = \int_{-\pi}^{\pi} \frac{d\phi}{2\pi}
\cos(b\phi) e^{\beta \cos(\phi) + \gamma \cos(2\phi)}
\label{eq:11}
\end{equation}
Here, it is convenient to solve eqs.(\ref{eq:10a},\ref{eq:10b}) in the
form $\beta(\beta_{V}),\gamma(\beta_{V})$ which we shall
call the ``Villain locus'' in the $\beta-\gamma$
plane. The comparison of our Monte Carlo data for the
internal energy of U(1)-LGT
in fig.4 clearly
demonstrates that the differences due to graphs of higher strength
$b = \pm 3,\ldots$
(which still have different weights) are completely
negligible - the data for the Villain action and the
mixed $\beta-\gamma$ action
(along the ``Villain locus'') fall almost
perfectly on top of each other.
  \begin{figure}
     \vspace{8.5cm}
     \caption[4]{The internal energy of U(1)-LGT with Villain
action. The lower data points are obtained from simulations
of the $\beta\cos\Theta$ model transformed according to
eq.(\ref{eq:9}). The
stars which fall practically on top of the Villain
data are from a simulation with mixed action
$\beta\cos\Theta + \gamma\cos 2\Theta$ treated according to
eqs.(\ref{eq:10a},\ref{eq:10b}).}
   \end{figure}

As a conclusion, the
differences between the Wilson and the Villain action
near the phase transition are completely
explained by the different weights of strength-2 ``stress''
graphs, whose proliferation is much more pronounced in the Wilson
case.

That an admixture of strength-2 graphs with large enough weight
can in principle drive the transition first-order, can be demonstrated
easily in the disorder field theory of ``stress'' lines (i.e. in the
mean-field formulation of the D=3 XY model with mixed action). This is
described by an effective action \cite{jkxy}
\begin{equation}
{\cal A} = a|\phi_{1}|^{2} + b|\phi_{1}|^{4} + c|\phi_{2}|^{2}
-\frac{1}{2} [ \phi_{1}^{2} \phi_{2}^{+} + c.c.] + {\rm gradients}
+ \ldots
\label{eq:last}
\end{equation}
where the complex fields $\phi_{1},\phi_{2}$ represent strength-1 and
strength-2 ``stress'' lines, respectively. The coupling $\phi_{1}
\phi_{1} \phi_{2}^{+}$ corresponds to the merging of two lines of strength-1
into one line of strength-2. The complex conjugate coupling describes
the reversed process. Integrating out the field $\phi_{2}$, one obtains
an additional quartic term in $\phi_{1}$, $-\frac{1}{4c} |\phi_{1}|^{4}$,
such that the total quartic term in $\phi_{1}$ may change sign, signalizing
a first-order transition.

This explains at least qualitatively why in careful
Monte Carlo studies with the Wilson action a
first-order transition was reported while with the Villain
action evidence for a continuous transition was claimed.
Moreover, this picture is consistent with Monte Carlo
studies of the mixed $\beta-\gamma$ action by
Jers\'{a}k et al.\cite{jersak}
who located a TCP for slightly negative $\gamma$.
Looking at fig.5, we see that our ``Villain locus'' indeed
crosses the phase transition line in the range
of second-order transitions. It would be interesting to
investigate whether the TCP is connected with a universal
tricritical ``2:1 ratio'' $\bar{W}_{2}/\bar{W}_{1}$.
   \begin{figure}
     \vspace{9.5cm}
     \caption[5]{The parameters $\beta,\gamma$ of the mixed
action $\beta\cos\Theta + \gamma \cos2\Theta$ which can be
studied by means of the improved Villain approximation
(``Villain locus''). The
fat and dashed lines show an interpolation to the critical
points found by Jers\'{a}k et al.\cite{jersak}. The dotted line was estimated
by those authors to be the locus of the Villain model in a
different way from ours (which we believe to be less
accurate).}
   \end{figure}

The most recent Monte Carlo renormalization group (MCRG) studies
are inconsistent with this nice picture. They are
interpreted as evidence for a first-order transition, with both
the Wilson and the Villain action. A historic summary is
compiled in table 3.
 \begin{table}
 \begin{center}
  \begin{tabular}{|l|c|c|c|}
   \hline 
   Author & $\gamma_{\rm TCP}$ & Wilson & Villain \\
   \hline\hline
   Bhanot (1982) \cite{bhanot}& $>0$ & 2 & 2 \\  \hline
   Jers\'{a}k et al. (1985) \cite{jersak}& $-0.11(5)$ & 1 & 2 \\  \hline
   Gr\"{o}sch et al. (1985) \cite{grosch}&       &   & metastability \\ \hline
   Morikawa et al. (1985) \cite{morikawa}& $>0$ & 2 & 2 \\ \hline
   Gupta et al. (1986) \cite{gupta}& $>0$ & 2 & 2 \\ \hline
   Burkitt (1986) \cite{burkitt}& $>0$ & 2 & 2 \\ \hline
   Lang (1986/87) \cite{lang}& $>0$ & 2 & -- \\ \hline
   Lang and Rebbi (1987) \cite{langreb}& -- & -- & 2 \\ \hline
   Decker et al. (1988) \cite{decker}& -- & 1 & -- \\ \hline
   A. Hasenfratz (1988) \cite{hase}& no & 1 & 1 \\ \hline
  \end{tabular}
 \end{center}
 \caption{Summary of work on the order of the phase 
transition in the D=4 U(1) lattice gauge model with Wilson and 
Villain action and on the location of the tricritical point 
in the parameter space of the mixed $\beta-\gamma$ action.} 
 \end{table}

The latest study \cite{hase} even speculates that the transition is always
first-order, downto $\gamma=-\infty$.
The last result would imply that in the whole parameter range of the D=4
Abelian Higgs model which is covered by the dual
$\beta-\gamma$ action, the Coleman-Weinberg mechanism is
indeed working. We feel, however, that even in view of this
impressive list of work in table 3, the
answer has not yet settled and much more high precision
studies are necessary to provide the final answer.

While in D=4 the situation is still quite
controversial, in D=3 dimensions the numerical evidence is
clearly against the analog of the Coleman-Weinberg
mechanism, as advanced by Halperin,Lubensky and Ma
\cite{halperin}. Here, it
was confirmed many times by different methods that both
Wilson's and Villain's action undergo continuous
transitions.
The
question was then whether there exists at all a parameter range in the mixed
$\beta-\gamma$ action which shows first-order transitions.
According to our above analysis, a good candidate was only
the range $\gamma \geq 0$. We therefore concentrated on this
range and found \cite{jkxy}, first in a mean-field (MF) treatment,
first-order
transitions for $0.166 \leq \gamma \leq 0.375$. The full
MF phase diagram is displayed in fig.6 (use the
labelings on the right and top axis).
  \begin{figure}
     \vspace{10.5cm}
     \caption[6]{Phase diagram of the D=3 XY model with mixed action as
obtained by mean-field methods and Monte Carlo simulations.}
   \end{figure}
Since already in the pure
XY model ($\beta = 0$ or $\gamma=0$ axis), the MF
transition temperatures are off by a factor
$0.45/0.33=1.36$, we have rescaled the MF curves by this
factor when comparing with our Monte Carlo simulations which
we run in the range where the rescaled MF results show the
largest entropy jump, $\gamma \approx 0.35...0.40$.
The numerical results confirmed the first-order nature of
the transition. As a typical signal, we show in fig.7 the
double peak structure in the internal energy histogram,
corresponding to tunnelings between two metastable states.
  \begin{figure}
    \vspace{6.0cm}
    \caption[7]{Energy histogram near the first-order
transition line of
the D=3 XY model with mixed action.}
  \end{figure}
Since the observed first-order transition is
very weak and the $\gamma$-range is probably very small, we
were not able to locate the two tricritical points with
resonable accuracy. We can therefore only claim evidence for
a short line of first-order transitions in the range $\gamma
\approx 0.35...0.40$.

In defect models of melting, no TCP seems to exist. The reason is the
very large activation energy of the graphs in the stress expansion
so that there are no pretransitional excitations up to the point
where the free energy intercepts the defect expansion. This is
responsible for a jump in the slope of the free energy, i.e. for
first-order transitions \cite{book}.

\end{document}